\documentclass[aps, reprint]{revtex4-2}
\usepackage[utf8]{inputenc}

\usepackage{bm}
\usepackage{tensor}
\usepackage{mathtools}
\usepackage{amssymb}
\usepackage{amsmath}
\usepackage{array}
\usepackage{enumitem}
\usepackage{listings}
\usepackage{booktabs}
\usepackage{graphicx}
\usepackage{subfig}
\usepackage{float}
\graphicspath{{./images/}}

\DeclareMathOperator*{\Cov}{Cov}
\DeclareMathOperator*{\CSP}{CSP}
\DeclareMathOperator*{\classify}{classify}

\begin{document}

\begin{abstract}
    Common Spatial Patterns (CSP) is a feature extraction algorithm widely used in Brain-Computer Interface (BCI) Systems for detecting Event-Related Potentials (ERPs) in multi-channel magneto/electroencephalography (MEG/EEG) time series data. In this article, we develop and apply a CSP algorithm to the problem of identifying whether a given epoch of multi-detector Gravitational Wave (GW) strains contains coalescenses. Paired with Signal Processing techniques and a Logistic Regression classifier, we find that our pipeline is correctly able to detect 76 out of 82 confident events from Gravitational Wave Transient Catalog, using H1 and L1 strains, with a classification score of $93.72 \pm 0.04\%$ using $10 \times 5$ cross validation. The false negative events were: \lstinline{GW170817-v3}, \lstinline{GW191219 163120-v1}, \lstinline{GW200115 042309-v2}, \lstinline{GW200210 092254-v1}, \lstinline{GW200220 061928-v1}, and \lstinline{GW200322 091133-v1}.
\end{abstract}

\title{Application of Common Spatial Patterns in Gravitational Waves Detection}
\author{Damodar Dahal}
\date{January 2022}

\maketitle

\section{Introduction}\label{intro}

Ever since the first LIGO run was conducted in 2015, 93 confident gravitational coalescenses have been identified \cite{GWTC-1, GWTC-2, GWTC-2.1, GWTC-3} till date by observing Gravitational Wave (GW) strains at one or more earth-based detectors. Such detections have presented several fruitful discoveries, including direct measurements of a black hole's spins \cite{BlackHoleSpin}, identification of neutron stars as the source of short gamma-ray-bursts \cite{ns-grb}, experimental confirmation of General Relativity \cite{testingGR}, and the birth of a novel field known as Multi-Messenger Astronomy (MMA) \cite{elipe2017multimessenger}. Various algorithms have been suggested and/or used in the GW search pipelines, including matched filtering \cite{matchedfilter1, GWTC-1, GWTC-2, GWTC-2.1, GWTC-3} and deep learning \cite{george2017deep, 2020dl}. In this article, we inherit an algorithm used in Brain-Computer Interface (BCI) systems, and use it for the first time to detect GWs.

The problem we are approaching is to identify whether $N$-detector Gravitational Wave (GW) strains, $\bm{\vec{h}}(t)$, consist of an Event-Related Potential (ERP) at a given time segment (or epoch) $[t, t+T]$ or not. Our events will be gravitational coalescenses, and we shall focus on the merger phase sinces these produce the highest potentials (amplitudes) in the detector strains. The core of our method uses Common Spatial Patterns (CSP), which produces a feature vector, $\bm{\vec{\phi}}$ of length $\leq N$ for any given epoch, which can be predict its class. It is widely used as a feature extraction algorithm in BCI \cite{CSP_lotte} to detect ERPs produced by mental arithmetic calculation \cite{CSP_MA} and visualization of imagined hand movements \cite{CSP_MI}. We shall assume that the strains are appropriately whitened and bandpassed beforehand, just like they are in BCI systems \cite{CSP_lotte}, correcting in our case for detector sensitivity, systemic and external noises. Our goal thus is to build and optimize the function,

\begin{equation}\label{problem}
    \hat y(t) = Y({\bm{\vec{h}}}(t)...\bm{\vec{h}}(t+T))
\end{equation}

corresponding to the probability of detecting an ERP between $[t, t+T]$. We sample the strains at $f_s$ Hz (eg: 4096 Hz = $2^{12}$ Hz), so the above equation can be written as:

\begin{equation}\label{main-eqn}
    \hat y(t) = Y(\bm{H}(t))
\end{equation}

where

\begin{equation}
    \bm{H}(t) \equiv \bigotimes_{i=0}^{f_sT}{\bm{\vec{h}}(t + i/f_s)}.
\end{equation}

Here, $\bm{H}(t)$ is a $(f_s T + 1, 0)$-rank tensor. We break $Y$ into intermediary ``feature extraction" and ``classification" steps:
\begin{equation} \label{eqn:y-break}
    Y \equiv \classify(\CSP(\bm{H}))
\end{equation}

$\CSP(\bm{H})$ represents the Common Spatial Patterns function, which reduces $\bm{H}$ into a ``feature vector" of length $0 < j \leq N$. We will discuss the methodology to find the function in the following subsection. For the classifier, we shall use Logistic Regression, although any other supervised learning algorithm like Support Vector Machines or Linear Discriminant Analysis can be used. Both $\CSP()$ and $\classify()$ have inherent ``hyperparameters" which shall be trained in due course.

\subsection{Common Spatial Patterns}

In Brain-Computer Interface (BCI) systems, Common Spatial Patterns has proven to be a robust feature extraction algorithm for detecting Event Related Potentials (ERPs) in one-vs-one classification problems \cite{CSP_lotte}, as well as in one-vs-many classification problems \cite{CSP_multiclass}. In one-vs-one problems, CSP produces the spatial filters, $\bm{W}$, such that it minimizes the variance of datasets of one class while maximizing the variance of the other class. CSP takes n-channel time series bandpassed data \cite{CSP_SIGFILT}; and since using GW strains from $N$-detectors form a similar $N$-channel time series data, we decided to try using CSP for feature extraction algorithm for GW detection. To date, no study has been performed using CSP on GW detection, and we show in Section \ref{results} that this feature extraction method produces successful results.

 We define our training set for developing the CSP spatial filters, $\bm{W}$, consists of $M$ datasets containing an ERP, $\{\bm{\mathcal{H}}_{\bm{+}, i}\}$, and another $M$ datasets which do not contain an ERP, $\{\bm{\mathcal{H}}_{\bm{-}, i}\}$. First, we estimate the covariances in each of the two classes, $\Cov(\bm{\mathcal{H}_{\pm}})$, and solve for simultaneous eigenvectors, which will become the bases for $\bm{W}$.

We calculate $\Cov(\bm{\mathcal{H}}_{\bm{\pm}, i})$, the covariance of each dataset, using shrinking covariance \cite{ledoit2004honey}. We then estimate the covariance of each class as the norm-traced mean of the covariance of individual datasets in the class, that is:

\begin{equation}
    \Cov(\bm{\mathcal{H}_{\pm}}) \equiv \frac{1}{|| \Cov(\bm{\mathcal{H}_{\pm}}) ||} {\frac{1}{M} {\sum_{i=0}^{M-1}{}\Cov(\bm{\mathcal{H}}_{\bm{\pm}, i})}}
\end{equation}

The norm trace is calculated using: 
$$||\bm{X}|| \equiv \sum_\mu{X_{\mu\mu}}$$

Now we search for simultaneous eigenvectors in the two covariance matrices such that:

\begin{equation}
    \Cov(\bm{\mathcal{H}}_+) \bm{\vec{v}}_{\bm{j}} = \lambda_j \Cov(\bm{\mathcal{H}}_-) \bm{\vec{v}}_{\bm{j}}
\end{equation}

where $0 \leq j < N$. This is a generalized eigenvalue problem with two symmetric matrices, which can be solved using numerical approximations \cite{ghojogh2019eigenvalue}. The CSP filter, $\bm{W}$, is made of the corresponding eigenvectors:

\begin{equation}
    \bm{W} = \bigotimes_{j=0}^{N-1}{\bm{\vec{v}}_{\bm{j}}}
\end{equation}

Finally, the components of our feature vector are given by:

\begin{equation}
    \bm{\phi}^{j} = \log{\left(\frac{1}{f_s T} \sum_{\kappa=0}^{N-1}{{\sum_{\nu=0}^{f_s T}{\bm{W}^{j \kappa} \bm{H}_{\kappa \nu}}}}\right)}, \quad 0 \leq j < N.
\end{equation}

Note that we are using contravariant and covariant components interchangeably.

\section{Methodology}

In the previous section, we described how two-class multi-channel time series data, one with ERPs, such as GW transients, and the other without them mapped are to feature vectors using Common Spatial Patterns which are separated from each other. We shall put that to use by detecting ERPs using 2-channel strain data from LIGO Hanford (H1) and LIGO Livingston (L1) detectors.

\subsection{Dataset}
Our dataset will include strains of two classes: i) ``event strains" consisting of GW transients and ii) ``baseline strains" sampled from random times within 4096s from each event strains. We shall use the strains for H1 (LIGO Hanford) and L1 (Ligo Livingston) detectors, and our events will be based on the confident detections of the Gravitational Wave Transient Catalog \cite{GWTC-1} \cite{GWTC-2} \cite{GWTC-2.1} \cite{GWTC-3}. Of the 93 available confident events, 11 of them had either missing or invalid data around (see Appendix \ref{blacklist}), so we excluded these transients from our dataset. As a result, below is the number of event transients comprising our dataset.

\begin{table}[htb]
    \centering
    \begin{tabular}{c | c}
        \toprule
                   Catalog &  Number of Events \\
        \midrule
          GWTC-1-confident &                11 \\
                    GWTC-2 &                31 \\
        GWTC-2.1-auxiliary &                 1 \\
        GWTC-2.1-confident &                 7 \\
          GWTC-3-confident &                32 \\
        \bottomrule
                     \textbf{Total} &        \textbf{82} \\
        
        \end{tabular}
    \caption{The catalogs we used to devise our dataset.}
    \label{tab:data-compos}
\end{table}

In total, we had $M=82$, or 82 event strain pairs and 82 baseline strain pairs for H1 and L1. These 4096 Hz strains were cropped to [-16s, +16s] around the event GPS time. The baseline strains were generated from the same data of the same length but of non-overlapping times.  For event strains, the GPS timestamp from the GWTC was centered. As a result, the dimensions of the dataset, $X$, and its labels, $y$, were $(164, 2, 131072)$ and $(164, 1)$ respectively.

Using Monte Carlo simulations of 10,000 runs, we compared pipelines with 3 different classifiers: Logistic Regression (LogReg), Support Vector Machine (SVM) and Linear Discriminant Analysis (LDA).

\section{Results}\label{results}
 The peak accuracy, validated using $10 \times 5$ cross-validation, was produced by Logistic Regression at $93.72 \pm 0.04 \%$ (mean $\pm$ standard deviation). It was produced through a four-step algorithm, which involved Window Selection (start = event GPS time - $0.06s$, length = $0.16s$, Signal Filtering (whitening + bandpass at [22, 272] Hz), Common Spatial Patterns (CSP), and one of the above classifier. The shrinkage regularization for estimating covariance matrix was set to 0.69. The LogReg classifier used $L_1$ penalty with regularization parameter, $C=0.56$, and it was iterated 708 times.
 
 We obtained the common CSP eigenvalues and eigenvectors:
\begin{align*}
\lambda_0 = 0.386507, \quad \bm{\vec{v}}_0 = (0.75836038, -1.96419165)\\
\lambda_1 = 0.393321, \quad \bm{\vec{v}}_1 = (-1.937204, -0.75598227) \end{align*}

Hence, the components of the CSP filter, $\bm{W} = \bm{\vec{v}}_0 \otimes \bm{\vec{v}}_1$, became:

\begin{equation}
    \bm{W}^{\mu \nu} = \left\lceil
    \begin{matrix}
    0.75836038 & -1.96419165\\
    -1.937204 & -0.75598227
    \end{matrix}
    \right\rceil
\end{equation}

\begin{figure*}
    \includegraphics[width=0.9\textwidth]{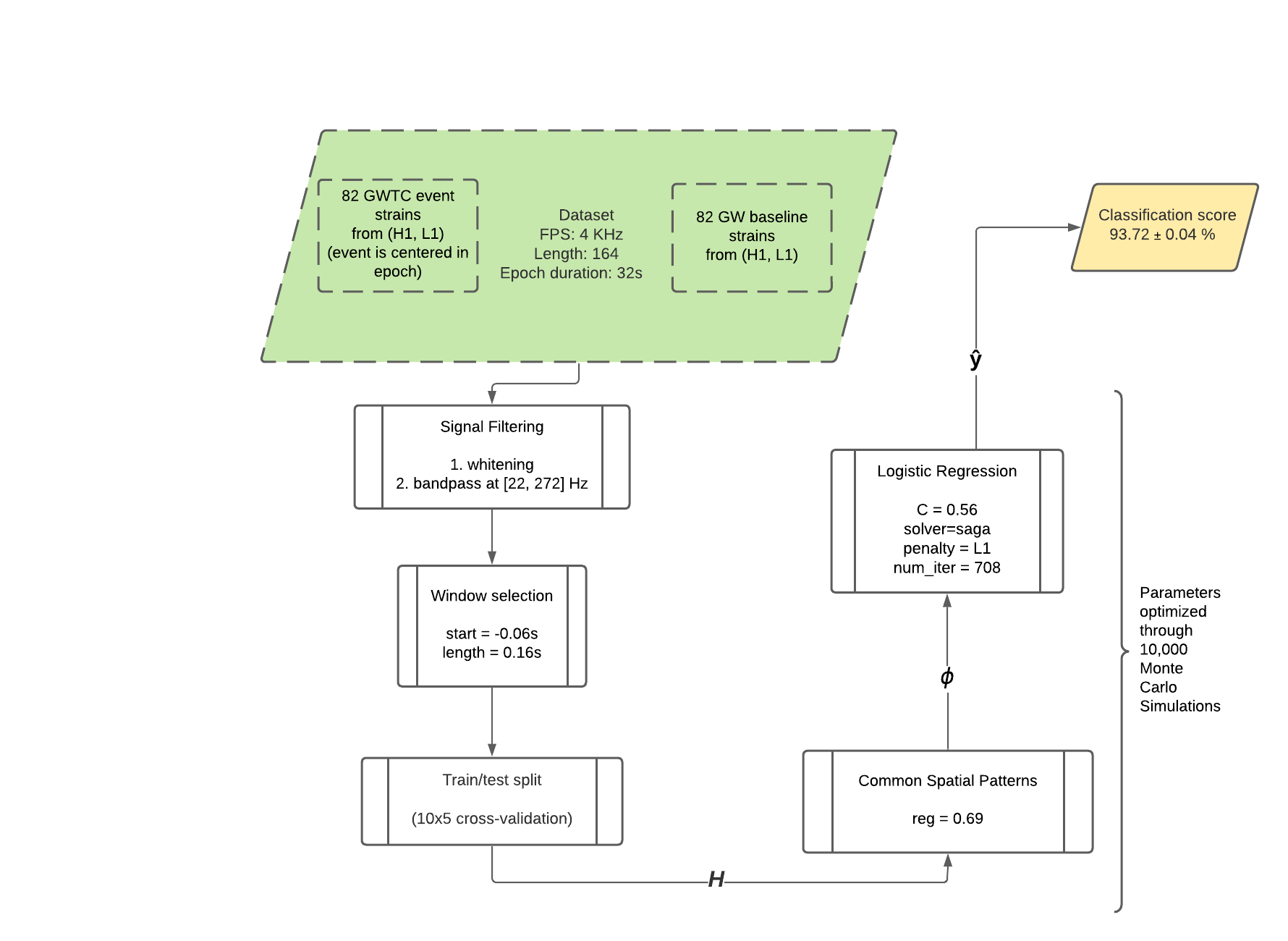}
    \caption{Our Supervised Learning Pipeline using CSP Feature Extraction}
    \label{fig:pipeline}
\end{figure*}

\begin{figure}[H]
    \centering
    \includegraphics[width=0.9\linewidth]{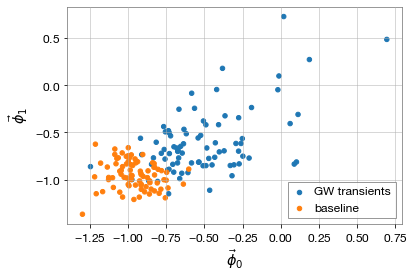}
    \caption{Visualizing features obtained through Common Spatial Patterns for (H1, L1) strain pairs. Each strain went through window selection, and a signal filter.}
    \label{fig:csp-vis-feature}
\end{figure}

In Figure \ref{fig:csp-vis-feature}, we plot the components of the feature vectors, $\bm{\vec{\phi}}$. Clearly, most events of each classes are grouped together, and the boundary line seems established even though some outliers are present.

\section{Evaluation}
\begin{figure}[H]
    \centering
    \subfloat[SNR of each event. The colors represent the accuracy of classifier in detecting the event. ]{\includegraphics[width=0.95\linewidth]{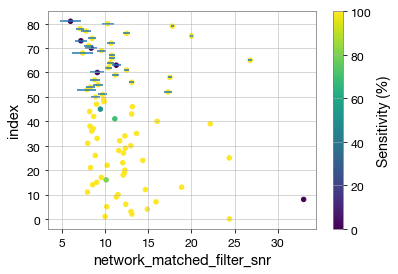}}\\
    
    \subfloat[Classifier sensitivity between 5dB $\leq$ SNR $\leq$ 33 dB.]{\includegraphics[width=0.9\linewidth]{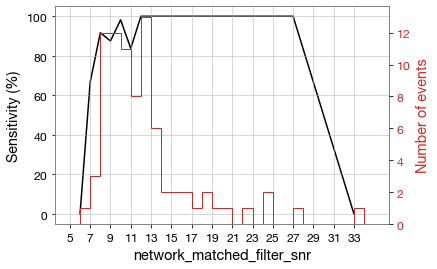}}
    \caption{Visualizing correlation classifier performance over $10 \times 5$ cross validation with the Signal-to-Noise (SNR) ratio provided by GW Transient Catalog. Due to the low frequency of events, Fig (b) may contain high error margins, especially for SNR $<$ 9 dB or $>$ 12 dB.}
    \label{fig:class-snr-perf}
\end{figure}

The pipeline was able to detect 76 of the 82 events with an average probability $\geq$ 0.5 during the $10 \times 5$ cross-validation runs. The events it failed to detect were \lstinline{GW170817-v3}, \lstinline{GW191219 163120-v1}, \lstinline{GW200115 042309-v2}, \lstinline{GW200210 092254-v1}, \lstinline{GW200220 061928-v1}, and \lstinline{GW200322 091133-v1}. For these false negative events, the greater source mass was between 1.46 and 34.00 \(M_\odot\), and the smaller source mass between 1.17 and 14.00 \(M_\odot\). Five of them had SNR between 6.0 and 11.3 dB (plotted in Fig \ref{fig:class-snr-perf}); however, GW170817-v3, the first detection of a neutron star collision, had a SNR of 33.0 dB, which was not detected by our algorithm. We suspect these issues lie in either the window selection or the frequency bands. It is known that these parameters impact the overall classification score (see Appendix \ref{param-impacts}), so our pipeline was optimized for maximum classification accuracy using one continuous frequency band and a specific window. In order to enable multiple frequencies and/or multiple windows, a method to run multiple CSP algorithms with each subset of data is described in \cite{CSP_lotte}, which will be studied on a separate work.

\medskip

\appendix

\section{Effect of Parameter Values of Classification Score}\label{param-impacts}
\begin{figure}[H]
    \centering
     \includegraphics[width=0.95\linewidth]{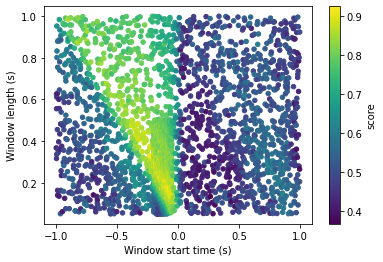}
    
    \caption{Visualizing classification scross across various window starts and duration. Each pipeline used whitening, bandpass at varied critical freqencies (lower: [10, 40] Hz, higher: [240, 370] Hz), CSP feature extraction, and Logistic Regression.}
    \label{fig:window-optimize}
\end{figure}

\begin{figure}[H]
    \centering
    \includegraphics[width=0.95\linewidth]{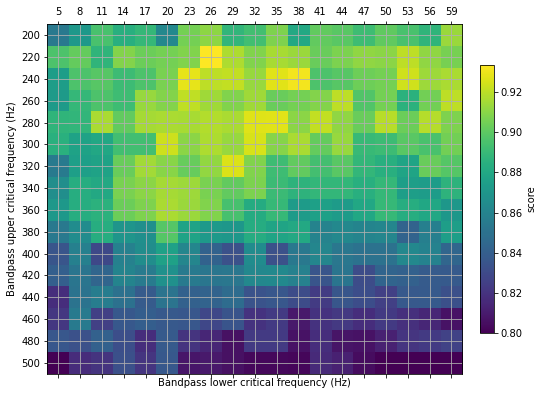}
    \caption{Visualizing maximum classification accuracy against a choice of critical bandpass frequencies. Each pipeline used whitening, bandpass, CSP feature extraction, and Logistic Regression.}
    \label{fig:sigfilt-optimize}
\end{figure}

We discuss the impact of some critical parameters during our 10,000 run Monte Carlo simulations. The information here might present useful in future works on GW detection with or without using CSP. In Fig. \ref{fig:window-optimize}, we show the impact of window selection on overall classification score, and in Fig. \ref{fig:sigfilt-optimize}, we show the impact of critical bandpass frequencies in the maximum classification scores.

\section{Blacklisted Events} \label{blacklist}
In this section, we list the events excluded from our dataset, either due to missing or invalid data.

\subsection{Missing Data}
Either the H1 or L1 strain covering [-16s, +16s] from event GPS timestamp was missing for the following events:

\begin{enumerate}[noitemsep]
    \item GW190421\_213856-v1
    \item GW190424\_180648-v2
    \item GW190620\_030421-v1
    \item GW190630\_185205-v1
    \item GW190708\_232457-v1
    \item GW190925\_232845-v1
    \item GW191216\_213338-v1
    \item GW200112\_155838-v1
    \item GW200302\_015811-v1
\end{enumerate}

\subsection{Invalid Data}
Either the H1 or the L1 strain of the following events around [-16s, +16s] from the event GPS time consisted invalid data (such as a \lstinline{NaN}).
\begin{enumerate}[noitemsep]
    \item GW190425\-v2
    \item GW190910\_112807-v1
\end{enumerate}

\bibliography{citations.bib}

\end{document}